# Superconductivity, What the H? The Emperor Has No Clothes

*Jorge E. Hirsch, Department of Physics, University of California San Diego*

A magnetic field *H* is expelled from the interior of a metal becoming superconducting [1]. Everybody thinks the phenomenon is perfectly well understood, particularly scientists with the highest *H*-index think that. I don't. I am convinced that without *H*oles, the little fiends that Werner *H*eisenberg conceptualized in 1931 [2], fifty years after Edwin *H*all had first detected them in some metals, you can't understand magnetic field expulsion nor anything else about superconductivity. Neither about the 'conventional superconductors' that are supposedly completely understood since 1957's BCS theory [3], nor about 'unconventional superconductors' like the high $T_c$ cuprates discovered in 1986, about which there is no agreement on anything *except* that they must be described by a *H*ubbard model [4]. I believe that this whole mess that we are in started with *H*erbert Fröhlich's [5] original sin [6] and the isotope effect experiments on *H*g [7] back in 1950, culminating in the current mania that metallic *H*ydrogen [8] or *H*ydrogen- rich alloys [9] will be (or already are! [10]) the first room temperature superconductors [11, 12]. I believe that the *H*ubbard model [13] has absolutely nothing to say about *H*igh temperature superconductivity nor any other superconductivity, despite the thousands of papers that have been written saying just that, and I believe that the theorem of *H*annes Alfven [14] is the key to understand the Meissner effect despite the fact that nobody else believes that.

There. In the above paragraph I tried to explain the title of this essay, why I have been a *H*eretic in the field of superconductivity for over 30 years, and why I believe that *H*ans' little story about the emperor [15] perfectly captures the essence of the situation. You don't have to believe any of it of course, it is certainly true that madness of crowds is far less probable than madness of an individual. In any event, here is (a highly condensed version of) the w*H*ole story [16].

For better or for worse, I am most famous (or infamous) for the invention of the *H*-index. I designed the *H*-index [17] to measure individual scientific achievement. It attempts to summarize the large amount of information contained in the number of citations to each of the papers you have written in a single number. Just in case you haven't yet heard about it, it is the number of papers that you have written that have more than that number of citations. If your *H*-index is 25, you have written 25 papers that each have 25 or more citations, the rest of your papers have fewer than 25 citations each.

I thought about this in 2003, tried it out for a couple of years, and in early 2005 wrote a preprint that I sent around to some colleagues but otherwise didn't know what to do with. A couple of months later, at the urging of Manuel Cardona, that had heard about it by word of mouth, I posted it on arXiv in early August of that year [18]. Manuel was a great physicist and a great human being, sadly deceased in 2014, that had a longstanding interest in bibliometrics, and had an extraordinarily high *H*-index.

The rest is history. The *H*-index has garnered wide attention, not only in physics but also in other natural sciences, social sciences, medicine, etc. Many papers have been written on its virtues, many more on its flaws, many variants of it have been proposed, yet so far none has been accepted as a better alternative.

In a nutshell, my observation is that about half the scientific community loves the *H*-index and half hates it, and the *H*-index of the scientist itself is a great predictor of whether s/he belongs to the first or the second group, in addition to its other virtues. I am not completely unhappy with the impact of my paper [19], which is by far my most highly cited one. As Oscar Wilde said, "There is only one thing in life worse than being talked about...".

I proposed the *H*-index hoping it would be an objective measure of scientific achievement. By and large, I think this is believed to be the case. But I have now come to believe that it can also fail spectacularly and have severe unintended negative consequences. I can understand how the sorcerer's apprentice must have felt.

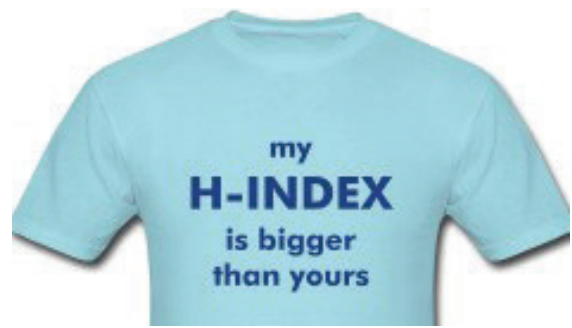

*Figure 1: Why there is no progress in understanding superconductivity.*

For example, if you are a student learning from your professor about the physics of superconductivity, and your professor is an expert in the field as proven by his/her high *H*-index, are you going to doubt that s/he understands the most basic physics of superconductivity? And knows the answers to the most elementary questions? Probably not. So you will listen carefully to what your professor tells you is well known and understood about superconductivity, put aside any qualms you might have based on your physical intuition and gut feeling, and not ask questions that sound too simple and may make you look stupid in the eyes of the professor. If those simple questions are not in the books, and the professor doesn't talk about them, they can't be valid questions. You will drink the Kool-Aid, learn how to work with the formalism, and later teach it to your students, who will be equally reluctant to question it as you were, since your $H$-index by then will be substantial.

The most highly recognized experts in the field of superconductivity have very high $H$-indices. They all agree unconditionally on some basic principles, namely: (*1*) *The BCS theory of superconductivity is one of the greatest, if not the greatest, achievement of modern condensed matter physics.* (*2*) *BCS is the correct theory to describe 'conventional superconductors', defined as materials described by BCS theory.* (*3*) *BCS is not the correct theory to describe 'unconventional superconductors', defined as materials that are not described by BCS theory.*

Wait, you will say–that's a tautology! True, so I should add: they also agree that the set of 'conventional superconductors' is not an empty set. And on that minor point I disagree. I am convinced that all superconductors are hole superconductors, and none is described by BCS theory.

Now my $H$-index is certainly astronomically smaller than the aggregate of the $H$-indices of all that are convinced that BCS theory is correct for conventional superconductors. It is also substantially smaller than that of many individuals that are highly recognized superconductivity experts in that group, e.g. Phil Anderson, Doug Scalapino, Marvin Cohen, Warren Pickett, Matthew Fisher, etc. Plus, the large majority of my papers that contribute to my $H$-index are not on the theory of hole superconductivity that I have been working on for the past 30 years. That work comprises about half of my total published work, and the aggregate citations to those 1,300 pages (which include a lot of self-citations) are less than 1/10 of my total citations, and less than 1/2 of the citations to my 4-page $H$-index paper [19].

So, if we believe citations and $H$-indices, by all counts my contributions to the understanding of superconductivity are insignificant.

Therefore, I have to conclude much to my regret that the $H$-index fails in this case. Because I know that the insights I have gained on hole superconductivity, in particular the realization that electron-hole asymmetry is the key to superconductivity, are far more important than any other work I have done that has a lot of citations, e.g. Monte Carlo simulating the (electron-hole symmetric) Hubbard model [20].

Already in early 1989 I was convinced that I had discovered a fundamental truth about superconductivity that nobody suspected: that only holes can give rise to superconductivity. As I wrote back then [21], "*the essential ingredient of our theory is the realization that holes are different from electrons...electrons in bonding states lead to attractive interactions between ions and repulsive interactions between electrons; electrons in antibonding states (holes) lead to repulsive interactions between ions and attractive interactions between electrons. The bonding electrons give lattice stability and normal metals, the antibonding electrons give lattice instabilities and superconductors...We expect this mechanism to account for the superconductivity observed in all solids.*"

When I excitedly told this to my senior colleague Brian Maple back then, I thought I had caught his attention. In the 70's Brian had worked closely with Bernd Matthias [22], the superconducting materials guru that had always been skeptical of BCS theory [23] and had often observed that lattice instabilities and superconductivity compete [24]. I asked Brian, after explaining why I thought that electron-hole asymmetry was the key to superconductivity and why this explained the connection between lattice instabilities and superconductivity that Matthias had obsessed about: "are you convinced?" I vividly remember his reply: "I'm convinced you are convinced".

I expected back then that 'this mechanism' would be quickly accepted by the community to be a self-evident truth, others would go on to develop it much further, and I could move on to work on other interesting topics. Alternatively, that somebody would prove me wrong, so I could move on. So where are we 30 years later?

I have not moved on. I have since then published well over 100 papers on hole superconductivity [16], going over many humps and hurdles to get around hostile referees, the papers have been by and large ignored and the community is as unconvinced as it was 30 years ago (or even more) that this has anything to do with real world superconductivity. Despite all the additional evidence I have found since 1989 that (in my view) strongly supports that my original conviction is right. Why is that?

One possible explanation is, of course, that I am wrong. The other more complicated explanation I believe is a combination of several factors: the opium of BCS theory [25], *H*-indices, paper-pushing grant managers and journal editors, lazy self-centered referees, and the emperor's new clothes [15].

Referees in particular. Did you ever have a hunch that referees are far more likely to view your paper favorably if it cites and/or talks favorably about their own papers? And that they are far less likely to give serious consideration to what your paper or grant proposal actually says and does or proposes to do if they get the impression that it undermines or potentially will undermine work that they have done? There is no remuneration for responsible refereeing nor is there a cost for irresponsible anonymous refereeing. Hence, given human nature and the refereeing system we have, it seems to me the guiding principle of refereeing in one sentence is: if publication of this paper or award of this grant will likely have a positive / negative effect on the *H*-index of the referee directly or indirectly in the future, s/he will recommend acceptance/rejection of the paper or grant proposal. Other criteria are second-order effects.

Ok, so what? Isn't that fair game, given that we are all both authors and referees? No, it is not to the extent that the game includes *others* that care that scientists that get paid by society to do work that supposedly ultimately benefits society actually do so.

Then there are the all important editors and grant managers. They decide who gets to referee your paper or grant proposal, typically go by rules of thumb that are the same for all papers and proposals, then do mindless vote-counting, oblivious to the difference between conforming papers or proposals and non-conforming ones. They have a natural tendency to pick referees that work on the same subject of your paper and don't take into account that if a paper questions the validity of a widely accepted theory such as BCS there is a conflict of interest with referees that have devoted their life and earned their reputation working with that theory.

BCS theory certainly made some valid points. Pairs undoubtedly play a role in superconductivity. Superconductors are macroscopically phase coherent. There is an energy gap in many superconductors.

But that can hardly justify the religious fervor with which the scientific community continues to cling to BCS theory today. One could understand it back in 1969, when Ron Parks compiled his famous treaty [26]. At that time, there was no reason to believe that more than one theory was needed to describe superconductivity in solids, and BCS was the only game in town.

But today? There are by a recent count [27] 32 different classes of superconducting materials, 12 of which are generally agreed to be 'conventional', i.e. described by BCS, 11 are generally agreed to be 'unconventional', i.e. not described by BCS, and 9 are 'undetermined', meaning there is no consensus whether they are BCS superconductors or not. So potentially 20 unconventional classes, where there is no agreement what is the mechanism governing them, versus 12 conventional, and we are still supposed to believe that BCS is the greatest achievement of modern condensed matter theory? Give me a break [28].

I believe that much of the explanation for this unconditional devotion to the conventional theory of superconductivity can be found in Andersen's little tale [15], that I will paraphrase here.

'*Many years ago there was an Emperor so exceedingly fond of new clothes that he spent all his money on being well dressed.*'

Many years ago there were physicists so enamored with their mathematical abilities to deal with complicated field theories that they forgot about physical reality.

'*clothes made of this cloth had a wonderful way of becoming invisible to anyone who was unfit for his office, or who was unusually stupid.*'

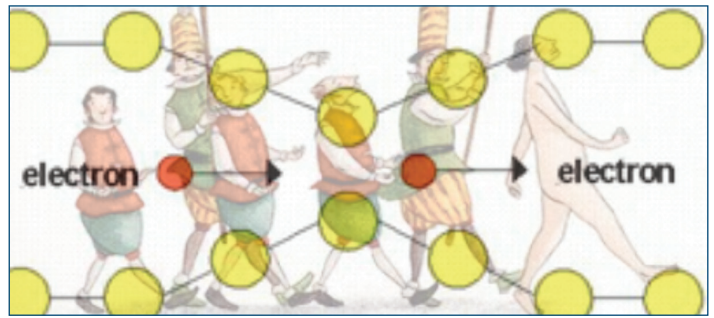

Figure 2: BCS: 'A theory of superconductivity is presented, based on the fact that the interaction between electrons resulting from virtual exchange of phonons is attractive when the energy difference between the electrons states involved is less than the phonon energy.' [3]

BCS-Eliashberg theory with the wonderful apparatus of field theory explains everything except to those that are unfit to be physicists or unusually stupid to comprehend it.

'*They set up two looms and pretended to weave, though there was nothing on the looms*'

They set out to predict the superconducting transition temperature of all the superconducting materials for which it had already been measured.

'*The whole town knew about the cloth's peculiar power, and all were impatient to find out how stupid their neighbors were.*'

All PRL referees knew about the peculiar power of BCS-Eliashberg-Bogoliubov-Ginsburg-Landau theory, and were impatient to reject the papers of stupid colleagues that would cast doubt on it.

*"'Heaven help me" he thought as his eyes flew wide open, "I can't see anything at all". But he did not say so.'*

"Heaven help me", thought smart students that couldn't understand how BCS theory explains the Meissner effect ."I can't possibly see how momentum conservation is accounted for and Faraday's law is not violated". But they did not say so.

*'They pointed to the empty looms, and the poor old minister stared as hard as he dared. He couldn't see anything, because there was nothing to see.'*

Theorists pointed to all the BCS calculations predicting new high temperature superconductors, and poor old experimentalists worked hard to make those superconductors and measure their $T_c$'s. They couldn't see anything, because there was nothing to see.

*' "I know I'm not stupid," the man thought, "so it must be that I'm unworthy of my good office. That's strange. I mustn't let anyone find it out, though". So he praised the material he did not see.'*

"I know I'm not stupid," experimentalists thought, "so it must be that I'm unworthy of my good office. That's strange. I mustn't let anyone find it out, though". And they wrote their papers explaining why their nonsuperconducting samples had made a mistake, and why the superconducting samples that they had found serendipitously perfectly matched BCS calculations.

*'The Emperor gave each of the swindlers a cross to wear in his buttonhole, and the title of "Sir Weaver." '*

The community awarded the theorists the Nobel prize, the Buckley prize, the Wolf prize, the John Bardeen prize, the APS medal, and membership in Academies and Royal Societies.

*'So off went the Emperor in procession under his splendid canopy. Everyone in the streets and the windows said, "Oh, how fine are the Emperor's new clothes! Don't they fit him to perfection?" '*

So off went the theorists to give talks, teach courses and write papers and books on their splendid theoretical framework. Everyone in the audiences, classrooms and reading rooms said, "Oh, how fine are these beautiful equations! Don't they fit observations on superconducting materials in the real world to perfection?"

*' "But he hasn't got anything on," a little child said..."But he hasn't got anything on!" the whole town cried out at last"... The Emperor shivered, for he suspected they were right. But he thought, " This procession has got to go on." So he walked more proudly than ever, as his noblemen held high the train that wasn't there at all.'*

"But these equations don't predict anything", Bernd Matthias said [29]... " "But they never have!", the whole experimental physics community cried out at last"... Senior theorists shivered, for they suspected they were right. But they thought, "This procession has got to go on." So they walked more proudly than ever, as their students, postdocs and junior collaborators held high the train that wasn't there at all.

And that is where we are today. The train isn't there at all. Let me explain why the emperor has no clothes.

Perhaps the simplest question you can ask about superconductivity is: how does a supercurrent stop? Even such a simple and fundamental question has never been asked, let alone answered, in the extensive literature on superconductivity. The answer is not trivial. When you heat a superconductor carrying a supercurrent across the superconducting transition, the current does not stop through onset of resistance. That would generate Joule heat, contradicting the fact that the transition is thermodynamically reversible.

Another fundamental question is: how does the Meissner effect work? Good conductors oppose changes in magnetic flux, and perfect conductors have magnetic flux lines frozen into them, because of Faraday's law. How come superconductors expel magnetic fields? How do they overcome Faraday's law, and satisfy momentum conservation? The final state carries a current that carriers momentum, the initial state does not. How does all that happen in a reversible way, without Joule heat dissipation, as required by thermodynamics?

Another related question: how does a rotating normal metal generate a magnetic field when cooled into the superconducting state? How do electrons defy inertia, some electrons spontaneously slowing down, others spontaneously speeding up, to generate the observed magnetic fields? How is angular momentum conserved?

Another question never asked before: when a superconductor in a magnetic field below $T_c$ is cooled further, how can the system reach a unique final state, independent of the rate of cooling, as BCS predicts, given that a variable amount of Joule heating is generated that depends on the speed of the process according to BCS theory?

These simple and fundamental questions have never been asked before in the BCS literature. There is nothing in BCS theory, the electron-phonon interaction, Cooper pairs, Bogoliubov quasiparticles, phase coherence, energy gap, spontaneous symmetry breaking, Higgs mechanism, Ginsburg-Landau theory, Eliashberg theory, that can say anything to answer the questions posed above.

I have asked these questions and showed that they can be answered within the theory of hole superconductivity, *if the normal state charge carriers are holes* [16].

Let me briefly explain the essential reason for it, it is simple and universal and can be explained in words. When you apply an external force to an electron near the bottom of the band, it aquires acceleration in the direction of the applied force, because its 'effective mass' is positive. We talk about holes rather than electrons when the Fermi level is near the top of the band. When you apply an external force to an electron near the top of the band, it aquires acceleration in direction opposite to that of the applied force, because its ef-fective mass is negative. What that means is simply that there is another force acting on the electron in opposite direction, that is larger than the applied external force. That other force originates in the coherent interaction of the electron with the periodic ionic lattice. In this situation then, there is *transfer of momentum* between the electrons and the body, while in the first case, when the electrons are near the bottom of the band, there isn't. This transfer of momentum occurs without scattering off impurities or phonons, so it does not generate entropy, it is a *reversible process*.

In superconductors, it is necessary to have a mechanism to *transfer momentum* between electrons and the body in a reversible way, to answer the questions listed above, how is momentum conserved when a supercurrent starts and stops. Therefore, holes are needed. Electrons cannot do it. It is as simple as that. The theory of hole superconductivity explains in detail how it happens [16].

It is generally believed that BCS theory predicts and explains the Meissner effect, but that is just not so. The BCS 'proof' of the Meissner effect [3] is a simple linear response argument, starting with the system in the BCS state and applying a magnetic field to it. That is *not* the Meissner effect. The Meissner effect is the process that starts with the system *in the normal state* with a magnetic field and ends up in the superconducting state with the magnetic field expelled. BCS theory says nothing about the process, other than the fact that the energy is lower in the final than in the initial state.

When I argue this with colleagues they will say, 'well you are talking about time dependence, sure, that is complicated, BCS correctly describes the equilibrium state though.' I answer, Faraday's law only acts if there is time dependence, so refusing to consider time dependence means refusing to acknowledge that Faraday's law exists and governs natural processes. If BCS theory does not have the physics that is necessary to explain how the system can go from the normal to the superconducting state expelling magnetic field against Faraday's law, it cannot be the correct theory of the equilibrium state either. Period.

The physics that explains how to expel magnetic fields, is quite simply, explained by Alfven's theorem [14]. A picture is worth a thousand words, the words are in my papers, the picture is at the right.

Alfven's theorem, on which the entire field of magnetohydrodynamics rests, is basically a restatement of Faraday's law. It states that in a perfectly conducting fluid magnetic field lines are frozen into the fluid and can only move together with the fluid. So why isn't it obvious that if magnetic field lines move out in the Meissner effect, it must be that fluid moves out, as in plasmas? Numquam ponenda est pluralitas sine necessitate. Of course there are several things to explain. What is the nature of the fluid that moves out, how can this happen without causing a charge and/or mass imbalance, what drives the motion, what does all this have to do with holes,

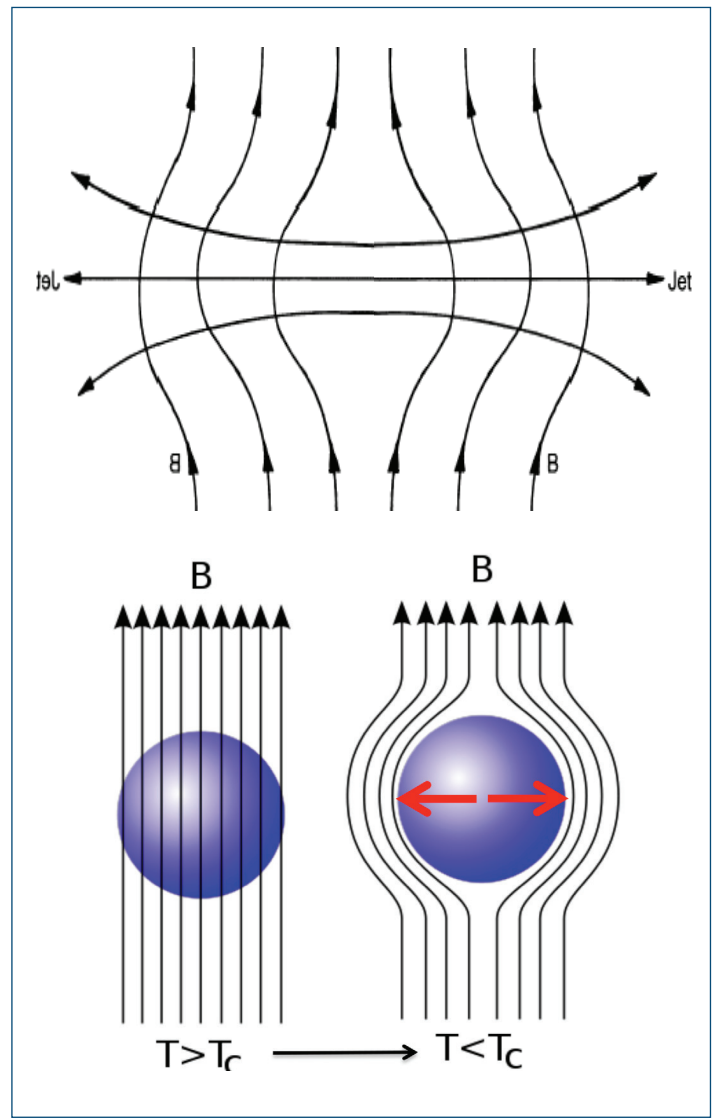

*Figure 3: Top panel: the right half is from a picture in ref. [14], with caption: 'An example of Alfven's theorem. Flow through a magnetic field causes the field lines to bow out.' I copied it, flipped the copy horizontally, and juxtaposed it to the left. On the bottom panel, a picture of the Meissner effect: as the temperature is lowered, it 'causes the field lines to bow out'. The red arrows show hypothesized 'Jets'*

etc. For all that, see references in the last 5 years in Ref. [16].

BCS theory does not describe fluid moving out, so it cannot describe the Meissner effect.

Many other reasons for why holes are necessary for superconductivity are given in the papers we wrote through the last 30 years [16], many in collaboration with my colleague Frank Marsiglio. Simple arguments show why $T_c$ is high in the cuprates and low in so-called 'conventional superconductors' [16], why there are 'electron-doped cuprates' [16], why the $T_c$ of $MgB_2$ is so high [16], why there is generically a positive isotope effect within this theory [16], even though the electron-phonon interaction has nothing to do with superconductivity, etc. Also, the periodic table shows a very significant correlation between sign of the Hall coefficient and superconductivity. Superconducting elements such as *Pb*, *Al*, *Sn*, *Nb*, *V*, *Hg*, etc, have positive Hall coefficient. Nonsuperconducting elements such as *Cu*, *Ag*, *Au*, *Na*, *K* have negative Hall coefficient. In 1997 I calculated that the probability that this is accidental, i.e. unrelated to superconductivity, is less than 1/100,000 [30].

For an overview of my work on hole superconductivity please see my recent book [31].

To sum up: either BCS is right, and then there is necessarily at the very least one other or more likely several other mechanism and physics of superconductivity, to describe the myriad of 'unconventional superconductors'. Or, BCS is wrong, and all superconductors are cut from the same cloth. If the latter can explain 140K superconductivity in the cuprates, it shouldn't have too much difficulty in accounting for the 7K superconductivity of Pb, should it? Yet Pb is held up as the 'posterchild' of BCS theory, that supposedly proves beyond reasonable doubt that only BCS can account for its existence.

No matter how hard I have tried, it has proven extraordinarily difficult to 'poke holes' in the BCS theory of superconductivity. Journal referees, grant managers, conference organizers, are extraordinarily resistant to allow consideration of heretic views on this topic, particularly in the US. Even sympathetic colleagues that profess to be open to the possibility that BCS may not be completely right are reluctant to undertake any serious consideration of the issues raised here, correctly assuming that it would undermine their chances to get their grants renewed, their salary raised, their invitation to speak at the next conference, and the growth of their *H*-index.

Maybe they are right, maybe they are not. If the latter, at some point the tide will turn, but it could be many years from now, when we are all gone. Meanwhile, as far as I can see, the emperor will continue to have no clothes. Which makes me unfit for my office and unusually stupid.